\documentclass[prd,eqsecnum,showpacs]{revtex4}
\usepackage{bm}
\usepackage{amssymb}
\begin{document}
\title{Universality of massive scalar field late-time tails in black-hole 
spacetimes}
\author{Lior M.~Burko}
\affiliation{Department of Physics, University of Utah, Salt Lake
City, Utah 84112}
\author{and Gaurav Khanna}
\affiliation{Physics Department, University of Massachusetts at
Dartmouth, North Dartmouth, Massachusetts 02747 \\ Natural Science
Division, Southampton College of Long Island University, Southampton, New 
York 11968} 
\date{\today} 
\begin{abstract}
The late-time tails of a massive scalar field in the spacetime of black
holes are studied numerically. Previous analytical results for a
Schwarzschild black hole are confirmed: The late-time behavior of the
field as recorded by a static observer is given by $\psi(t)\sim
t^{-5/6}\sin [\omega (t)\times t]$, where $\omega(t)$ depends weakly on
time. This result is carried over to the case of a Kerr black hole. In
particular, it is found that the power-law index of $-5/6$ depends on
neither the multipole mode $\ell$ nor on the spin rate of the black hole
$a/M$. In all black hole spacetimes, massive scalar fields have the same
late-time behavior irrespective of their initial data (i.e., angular
distribution). Their late-time behavior is universal.
\end{abstract}
\maketitle

\section{Introduction}

The late-time behavior of massless fields in black hole spacetime has been
studied in detail for both linearized \cite{price-72,burko-khanna} and 
fully nonlinear (spherical) evolutions \cite{GPP2,BO}. In contrast, the
late-time behavior of massive fields has been studied in much less
detail. In this paper we study the late-time behavior of a scalar field
numerically, in the spacetimes of Schwarzschild and Kerr black holes. 

The first to consider the problem of a massive scalar field in the
spacetime of a Schwarzschild black hole were Novikov and
Starobinski, who studied the problem in the frequency domain, and found
that there are poles in the complex plane closer to the real axis than in
the massless case. They thus inferred that the decay rate would be slower 
in the massive case than in the massless case \cite{starobinski}. That
problem was later studied numerically by Burko \cite{burko-97} both for a 
linearized massive scalar field, and for a self-gravitating, 
spherical massive field. In both cases it was found in \cite{burko-97}
that the late-time behavior was given by an oscillating field, whose
envelope decayed according to an inverse power law, and whose frequency
was determined by the Compton wavelength of the field, i.e., by its mass
term. In \cite{burko-97} it was also reported that the decay rate of the
envelope was given by $t^{-\alpha}$, where $\alpha\sim 0.8$. No attempt
was done in \cite{burko-97}, however, to determine the value of $\alpha$
accurately, or to determine its dependence or lack thereof of the
parameters of the problem. The most detailed analytical study of the
problem was done by Koyama and Tomimatsu in a series of three papers
\cite{koyama1,koyama2,koyama3}. 

The most striking feature of the problem of tails of a massive scalar
field is that the tails exist already in flat spacetime \cite{morse},
because spacetime acts like a dispersive medium for a Klein-Gordon
field. Based on the exact Green function, the 
behavior of the field in Minkowski spacetime was found in \cite{burko-97}
to be given by
\begin{equation}\label{flat}
\psi_{\rm flat}\sim t^{-\ell-3/2}\sin(\mu t)\, ,
\end{equation}
$\ell$ being the multipole moment of the field, and $\mu ^{-1}$ being its 
Compton wavelength. A derivation of Eq.~(\ref{flat}) appears in the
Appendix. More recently, an interesting self similar behavior of a 
Klein-Gordon field in Minkowski spacetime was found in
\cite{fodor_racz_03}. The self similar behavior reduces to
Eq.~(\ref{flat}) for a field evaluated at a fixed location at very late times. 

In the spacetime of a Schwarzschild black hole there
are important differences. The decay rate was found in \cite{koyama2} to
be given by 
\begin{equation}
\psi_{\rm sch}\sim t^{-5/6}\sin[\omega (t)\times t]\, ,
\end{equation} 
where the power index is independent of $\ell$. The time-evolving angular
frequency $\omega (t)$ approaches $\mu$ asymptotically as $t\to\infty$,
but is different from $\mu$ at finite values of time. This result was
generalized in \cite{koyama3} to 
any spherically symmetric, static black hole spacetime. The details of
this result have not been corroborated numerically---although
qualitatively
they are supported by the results of \cite{burko-97}---a task we
undertake in this paper. In addition, we also show that this result
remains correct also for a Kerr black hole, for which the power index
$-5/6$ is independent of both $\ell$ and the spin rate of the black hole
$a/M$. We also study the behavior of $\omega(t)$, and find it to be in
agreement with the results of \cite{koyama2,koyama3}. Similar conclusions
were obtained recently also for a spinning dilaton black hole
\cite{pan_jing} (see also \cite{moderski_rogatko}).

\section{Massive tails in Schwarzschild spacetime}

Massive fields in Schwarzschild spacetime were studied numerically (for
both a linearized and for a spherical self-gravitating Klein-Gordon field) 
in \cite{burko-97} and analytically in \cite{koyama2}. As discussed in
\cite{koyama2}, for $\mu M\lesssim 1$, the tail regime satisfies $t\gg
\mu^{-3}M^{-2}$, $M$ being the mass of the black hole, and $t$ 
being the regular Schwarzschild time coordinate. Therefore, for a
light scalar field, the heavier the field is, the earlier the tails
are seen. (For heavy fields, $\mu M\gg 1$, the tail regime is for $t\gg
M$, independently of the mass of the field.) To facilitate numerical
simulations, one is tempted then to
increase the mass of the field. However, increasing the mass of the scalar
field requires also to increase the resolution, as the wavelength of the
field is inversely proportional to its mass, and as one needs at least a
few data points per wavelength to resolve an oscillating field. One faces
then a competing effect: to decrease the computational time to an optimum
one needs to balance the scalar field mass and the resolution of the
numerical code. (We do not discuss here the case of $\mu M\gg 1$ as the
required numerical resolution is beyond current practical computational
limits.) One may gain insight into the above discussion, noticing,
following \cite{burko-97}, that for $\mu\ll 1/M$, the
corresponding Compton wavelength of the scalar field   
$\lambda(=\mu^{-1})\gg M$, that is the field's wavelength is much longer
than the typical radius of curvature of spacetime, or the scale of
inhomogeneity of curvature. We therefore expect the field at early times
to evolve similarly to its evolution in flat spacetime. At later times the
curvature effects become apparent, and we expect deviations from the
flat-spacetime behavior. At asymptotically late times we expect to find the
behavior predicted in \cite{koyama2}. The greater the mass term $\mu$, the
shorter the associated Compton wavelength $\lambda$, and the sooner the
asymptotic domain is expected. We therefore expect an oscillating field,
whose envelope is a broken power law: at early times it is described by
$t^{-\ell -3/2}$ and at late times by $t^{-5/6}$.

The field $\psi$ satisfies the Klein-Gordon equation 
\begin{equation}
(\Box-\mu^2)\phi=0\, ,
\end{equation}
$\Box$ being the D'Alembertian operator in curved spacetime. We define
$\psi=r\phi$ in the usual way, and decompose the field $\psi$ into
Legendre modes $\psi=\sum_{\ell}\psi_{\ell}P_{\ell}(\cos\theta )$. The
radial equation that each mode $\psi_{\ell}$ satisfies is then given by
\begin{equation}
\psi^{\ell}_{,uv}+\frac{1}{4}V_{\ell}(r)\psi^{\ell}=0\, ,
\end{equation}
where the effective potential 
\begin{equation}
V_{\ell}(r)=\left(1-\frac{2M}{r}\right)\left[\frac{\ell (\ell +1)}{r^2}
+\frac{2M}{r^3}+\mu^2\right]\, .
\end{equation}

To test our expectations in the context of a Schwarzschild black hole,
we used a double-null code in 1+1D, and solved the radial equation for 
$\psi^{\ell}$. We specified a pulse of the form 
\begin{equation}
\psi^{\ell} (u)=\left[4\frac{(u-u_1)(u-u_2)}{(u_2-u_1)^2}\right]^8
\end{equation}
along the ingoing segment of the characteristic hypersurface $v=0$ 
for $u_1<u<u_2$, and $\psi (u)=0$ otherwise. We also took $\psi (v)=0$
along $u=0$. Here, $u$ ($v$) is the
usual retarded (advanced) time.  
The results reported on below correspond to $u_1=20M$ and $u_2=60M$, but
our results are unchanged also for other choices of the parameters, or for
other choices of the characteristic pulse. In the following we observe the
fields along $r_*=0$, $r_*$ being the usual Regge-Wheeler `tortoise'
coordinate. 

\begin{figure}
\input epsf
\centerline{ \epsfxsize 8.0cm
\epsfbox{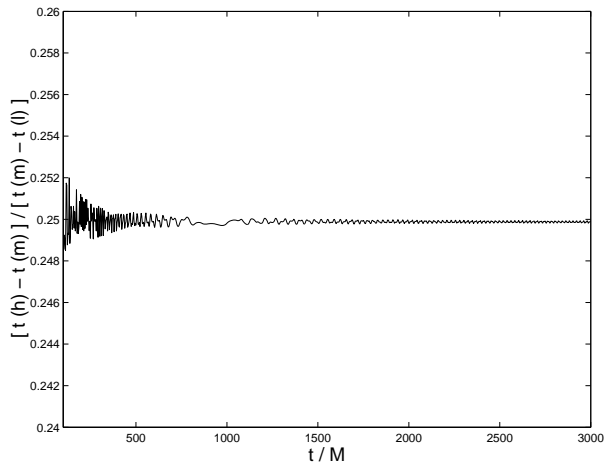}}
\caption{Convergence test for the 1+1D code. We find the times at which
the field has zeros for three different resolutions, high $t({\rm h})$,
medium $t({\rm m})$, and low $t({\rm l})$, and compute the ratio
$[t({\rm h})-t({\rm m})]/[t({\rm m})-t({\rm l})]$ for each zero. We then
plot this ratio as a function of time $t$. Here, we
used $N=5,10$, and 20 grid points per $M$ for the three
resolutions. Second order convergence
corresponds to the ratio equaling 0.25.  }
\label{conv_sch}
\end{figure}

The code is stable, and exhibits second order convergence. The convergence
test is presented in Fig.~\ref{conv_sch}, which shows the ratio of
differences for three runs with different resolutions.

\begin{figure}
\input epsf
\centerline{ \epsfxsize 8.0cm
\epsfbox{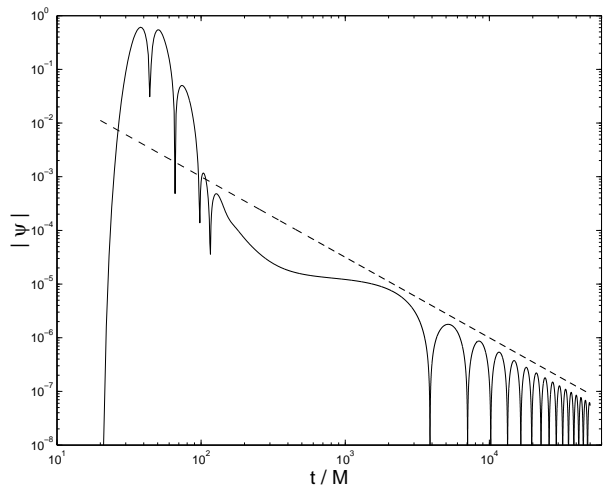}}
\caption{Field of a massive scalar field along $r_*=0$. The initial data
are for $\ell=0$, $u_1=20M$ and $u_2=60M$, and $N=1$. The solid curve is
the massive scalar field (with $\mu M=10^{-3}$), and the dashed line is
proportional to $(t/M)^{-3/2}$. }
\label{low_mu}
\end{figure}

We first chose a low value of $\mu M$. Figure \ref{low_mu} shows the
field along $r_*={\rm const}$ as a function of time. The initial field
here is spherical ($\ell=0$), and a value of $\mu M=10^{-3}$ was
chosen. Similar results are obtained also for other low values of $\mu M$.
After the prompt burst and the quasi-normal epoch, an oscillatory tail is
seen. The decay rate of the envelope of the tail is very close to
$t^{-3/2}$. 

\begin{figure}
\input epsf
\epsfxsize=8.5cm
\centerline{ \epsfxsize 8.0cm
\epsfbox{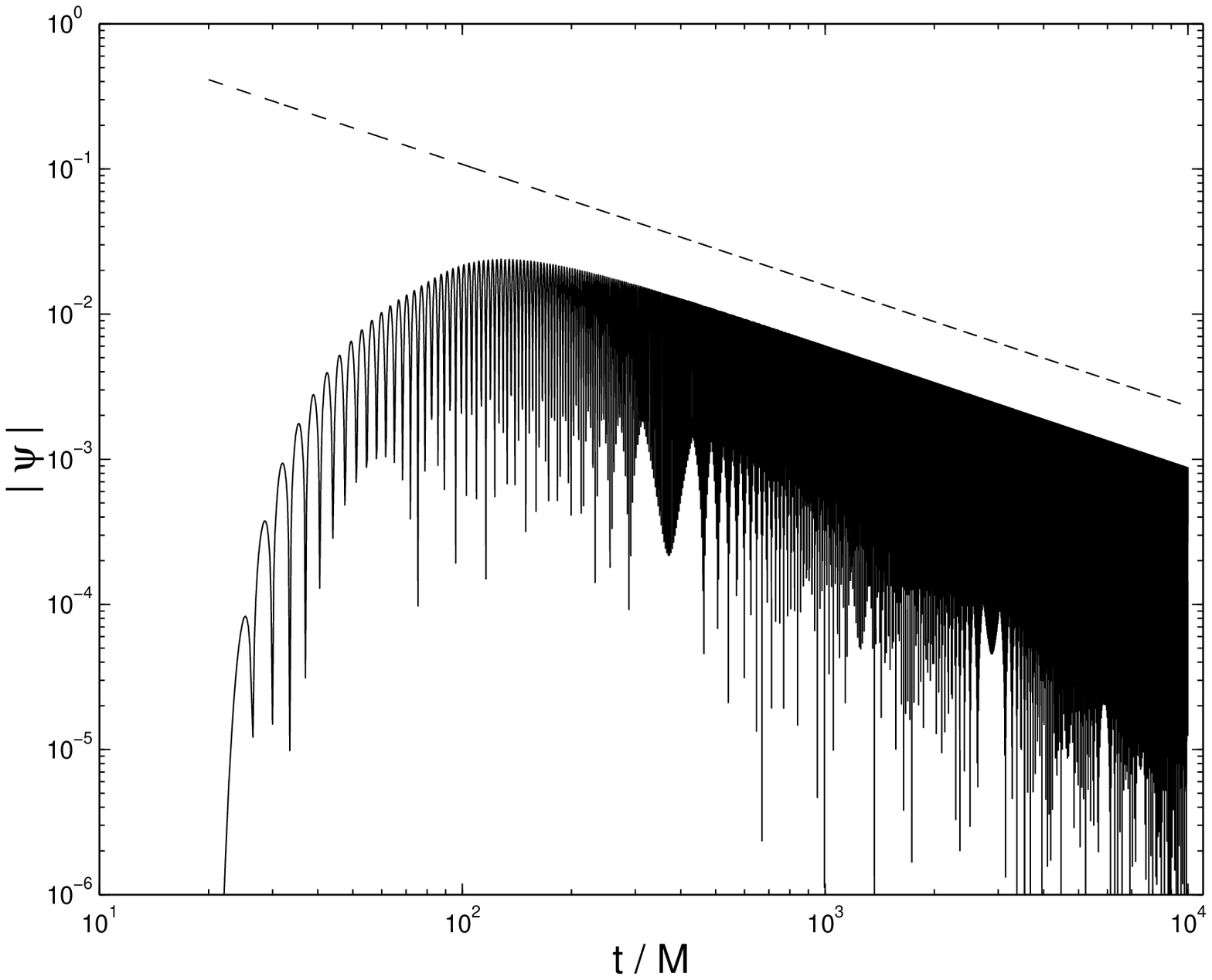}}
\caption{Field of a massive scalar field along $r_*=0$. The initial data
are for $\ell=0$, $u_1=20M$ and $u_2=60M$, and $N=4$. The solid curve is
the massive scalar field (with $\mu M=1$), and the dashed line is
proportional to $(t/M)^{-5/6}$. }
\label{high_mu}
\end{figure}

We next consider a high value of $\mu M$. Figure \ref{high_mu} shows the
field for the same parameters as in Fig.~\ref{low_mu}, except that here 
$\mu M=1$. Again, an oscillatory tail is seen, and the decay rate of the
tail is very close to $t^{-5/6}$.

\begin{figure}
\input epsf
\centerline{ \epsfxsize 8.0cm
\epsfbox{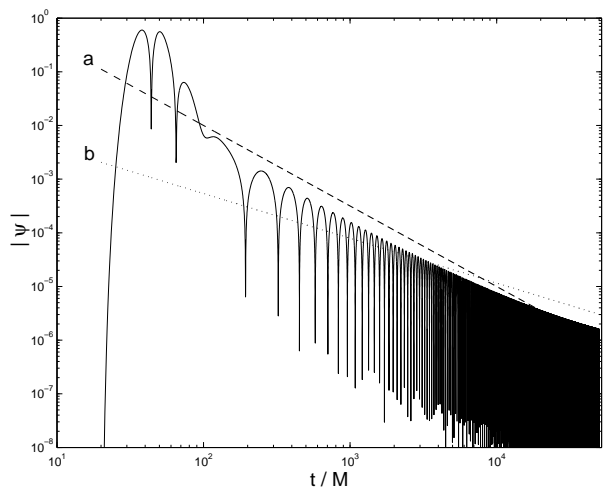}}
\caption{Field of a massive scalar field along $r_*=0$. The initial data
are for $\ell=0$, $u_1=20M$ and $u_2=60M$, and $N=4$. The solid curve is
the massive scalar field (with $\mu M=2.5\times 10^{-2}$). The dashed line
(a) is proportional to $(t/M)^{-3/2}$, and the dotted line (b) is
proportional to $(t/M)^{-5/6}$. }
\label{medium_mu}
\end{figure}

The most interesting case is that of an intermediate value of $\mu M$, for
which both types of behavior coexist. In Figure \ref{medium_mu} we show
the field for $\mu M=2.5\times 10^{-2}$. This figure shows initial
oscillatory decay with envelope decaying like $t^{-3/2}$, which changes
gradually to an oscillatory decay with decay rate of $t^{-5/6}$. This is
the broken power-law expected: at early times the field has not noticed
yet the presence of the black hole, and therefore behaves as in flat
spacetime (decay rate of $t^{-3/2})$. At late times, the field takes its 
asymptotic decay rate of $t^{-5/6}$, and the onset of the asymptotic
behavior can be controlled by changing the value of $\mu M$.

\begin{figure}
\input epsf
\epsfxsize=8.5cm
\centerline{ \epsfxsize 8.0cm
\epsfbox{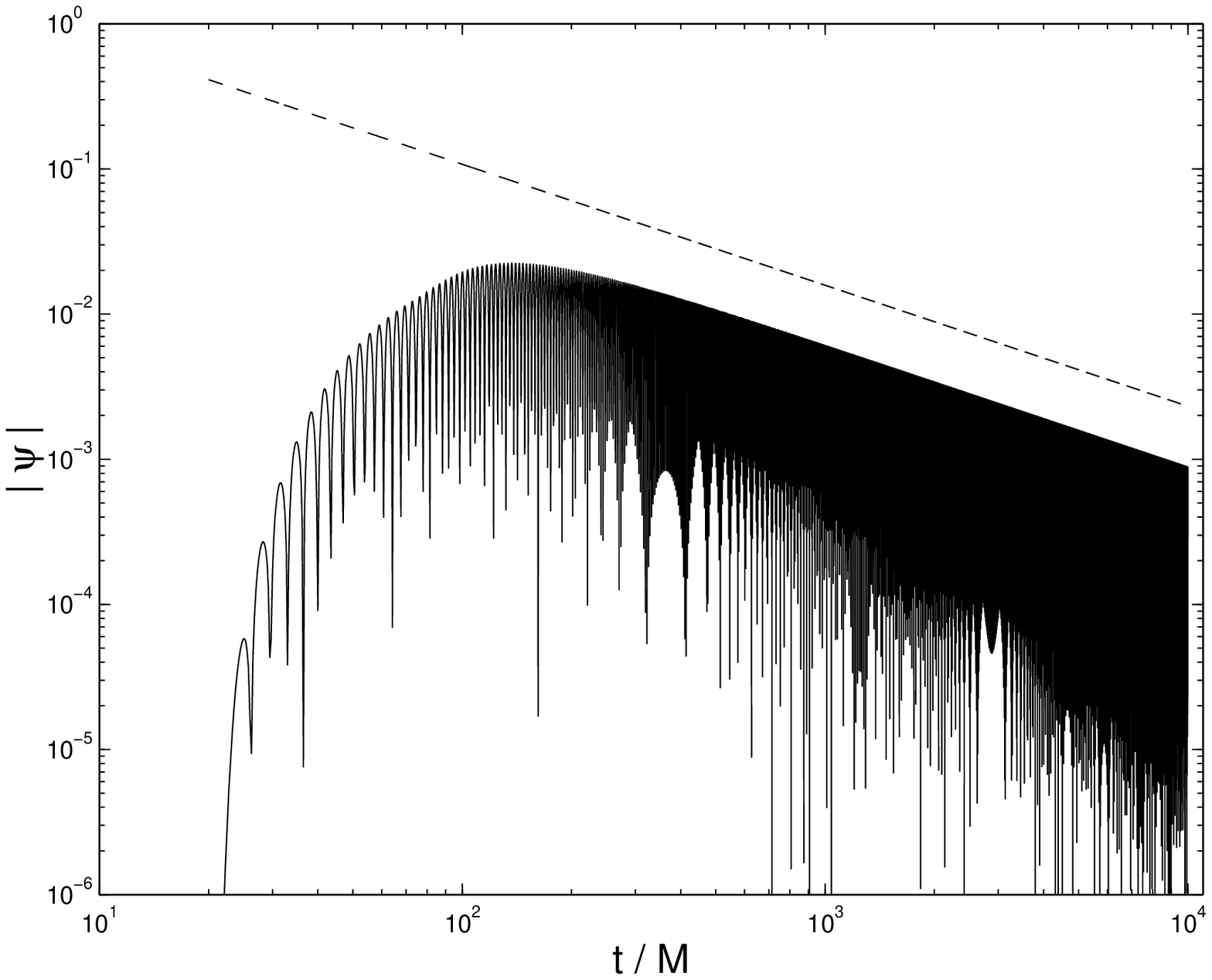}}   
\caption{Same as Fig.~\ref{high_mu}, but with $\ell=1$.}
\label{ell_1}
\end{figure}

The prediction of Ref.~\cite{koyama2} is that the asymptotic decay rate of
$t^{-5/6}$ is independent of the value of $\ell$. We check this in
Fig.~\ref{ell_1} which shows that same data as in Fig.~\ref{high_mu},
except that here we take $\ell=1$. Again, the decay rate is $t^{-5/6}$,
the same as with $\ell=0$. We could not find any deviation from the
$t^{-5/6}$ behavior for any value of $\ell$. We therefore conclude that
our simulations are in agreement with the prediction of Ref.~\cite{koyama2}.

In flat spacetime the period of the oscillations does not change at
late times, and is given by $T_{\rm flat}=2\pi/\mu$ (see the Appendix). In
Schwarzschild spacetime, the period is no longer fixed. In
Ref.~\cite{koyama2}, the oscillatory part of the field is given by 
$$\sin \left[\mu t -(3/2)(2\pi \mu M)^{2/3}(\mu t)^{1/3}+ {\rm
smaller\;\; terms}\right]\, .$$
To interpret this prediction as a changing period, we re-write the
oscillatory part of the field as $\sin [\omega(t)\times t]$, where 
\begin{equation}
\omega(t)=\mu \left[ 1-\frac{3}{2}\left(2\pi\frac{M}{t}\right)^{2/3}+
O\left(\frac{1}{t}\right)\right]\, .
\end{equation}

\begin{figure}
\input epsf
\centerline{ \epsfxsize 8.0cm
\epsfbox{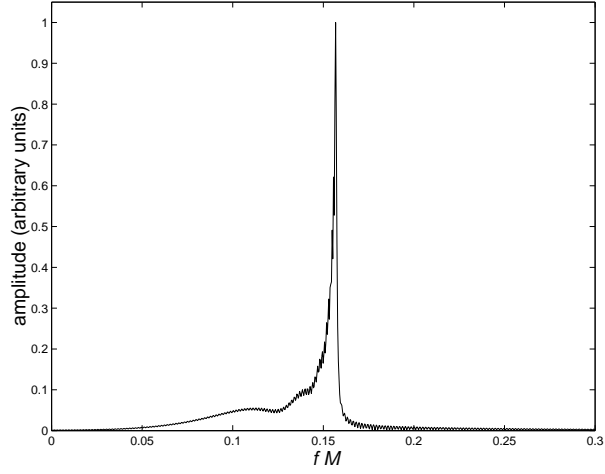}}
\caption{Power spectrum of the field. Here, $\mu M=1$, and $N=100$.
To obtain this plot, the field's values were first scaled by $t^{5/6}$ 
to get close to a pure sinusoid, and then the Fourier transform was 
taken. We assume here (and below) $M=1$.}
\label{fft_sch}
\end{figure}

\begin{figure}
\input epsf
\centerline{ \epsfxsize 8.0cm
\epsfbox{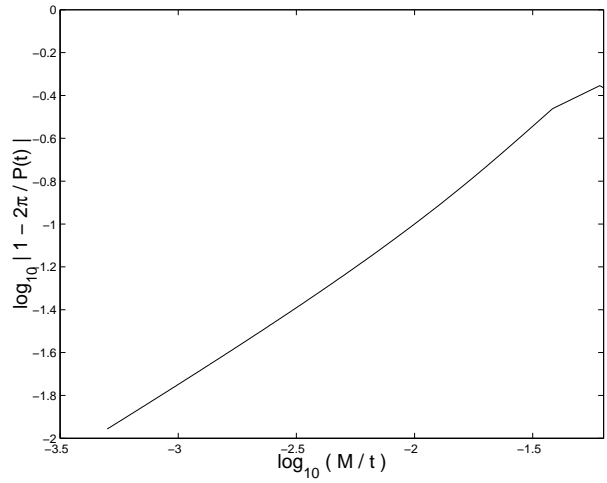}}
\caption{The changing period: $1-2\pi / P(t)$ vs.~$M/t$, for the same
data as in Fig.~\ref{fft_sch}.}
\label{p_of_t}
\end{figure}

\begin{figure}
\input epsf
\centerline{ \epsfxsize 8.0cm
\epsfbox{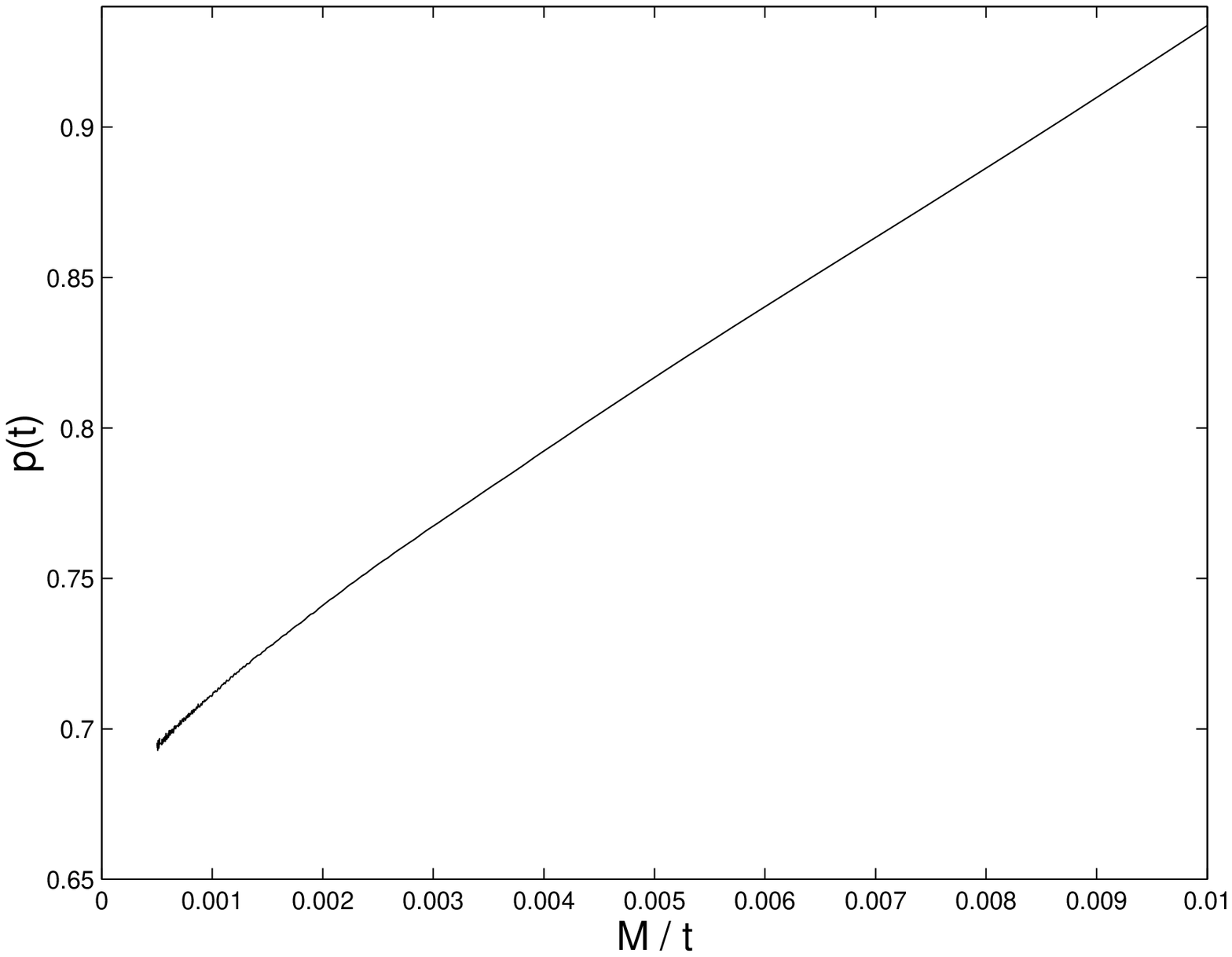}}
\caption{The local slope of the curve in Fig.~\ref{p_of_t}, $p(t)$, as a 
function of $M/t$.}
\label{local_slope}
\end{figure}

Figure \ref{fft_sch} shows the power spectrum of the field, for the
case $\mu M=1$. A sharp peak at $f=(2\pi )^{-1}$ is seen. This peak
corresponds to angular frequency equaling the mass term $\mu$. Notice the
one-sided broadening of this peak. While no frequencies higher than $(2\pi
)^{-1}$ appear to be present, lower frequencies are. This suggest that as
time progresses, the frequency increases to its asymptotic value of 
$(2\pi )^{-1}$. We test this in Fig.~\ref{p_of_t} by plotting, $1-2\pi / P(t)$ 
as a function of $t^{-1}$. Here, $P(t)$ is the local
period of the oscillations. We find that as $t\to\infty$, indeed the
period decreases to its asymptotic value. Next, we consider in greater
detail the rate at which the period approaches its asymptotic
value. Recall that according to Ref.~\cite{koyama2}, the slope of the
curve in Fig.~\ref{p_of_t} should be $2/3$. In Fig.~\ref{local_slope} we
display the local slope of the curve as a function of $t^{-1}$. Our
numerical result, of an asymptotic slope of $0.66\pm 0.01$, is in
agreement with the predictions of Ref.~\cite{koyama2}.

\section{Massive tails in Kerr spacetime}

Now we turn to the discussion of massive scalar field tails in Kerr 
background. We intuitively expect the late-time behavior in this case
to be identical to that of the Schwarzschild case, as presented in the 
previous section. The reason why the late-time tail is expected to be  
independent of  $a/M$ is the following: For a massless field in Kerr 
spacetime, there is a mixing of modes and each existing mode decays 
with a decay rate of $t^{-(2\ell+3)}$, which is the same decay rate as in
the Schwarzschild case. All the modes which are not disallowed are
excited, such that the overall decay rate is dominated by the existing
mode with the slowest decay rate \cite{burko-khanna}.  We expect this
situation to remain basically the same also for a massive field. 
However, for a massive scalar field in Schwarzschild spacetime the decay
rate is independent of the mode $\ell$. This implies that although in
Kerr mode mixing will indeed generate more modes and each mode will have
the same evolution as in Schwarzschild spacetime (as in the massless
case), because in the Schwarzschild case all massive field modes have
the same decay rate, they will all have the same decay rate in Kerr
spacetime. 

Our numerical simulations were performed using the penetrating 
Teukolsky code (PTC) \cite{ptc}, which solves the Teukolsky equation for 
linearized perturbations over a Kerr background in the ingoing Kerr 
coordinates $({\tilde t},r,\theta, {\tilde \varphi})$. These coordinates are 
related to the usual Boyer-Lindquist coordinates $(t,r,\theta,\varphi)$ through 
${\tilde \varphi}=\varphi+\int a\Delta^{-1}\,dr$ and 
${\tilde t}=t-r+r_*$, where $\Delta=r^2-2Mr+a^2$ 
and $r_*=\int(r^2+a^2)\Delta^{-1}\,dr$. The Teukolsky equation for the 
function $\psi$ in the ingoing Kerr coordinates can be obtained by 
implementing black hole perturbation theory (with a minor rescaling of the 
Kinnersley tetrad \cite{ptc}). It has no singularities at the event horizon, and
therefore is capable of evolving data across it. This equation is given
for a massless field ($\mu=0$) by 
\begin{eqnarray}\label{teukolsky}
&&
(\Sigma + 2Mr){{\partial^2 \psi}\over
{\partial \tilde t^2}} - \Delta {{\partial^2 \psi}\over
{\partial r^2}} + 2(s-1)(r - M){{\partial \psi}\over
{\partial r}} \nonumber
\\
&&
-{{1}\over {\sin \theta}}{{\partial}\over {\partial \theta}} \left (   
\sin \theta {{\partial \psi}\over {\partial \theta}}\right ) -{{1}\over
{\sin^2 \theta}}{{\partial^2 \psi}\over {\partial \tilde \varphi^2}}  
-4Mr{{\partial^2 \psi}\over {\partial \tilde t \partial r}}\nonumber\\
&&
-2a {{\partial^2 \psi}\over {\partial r\partial \tilde \varphi}}
- i {2s\cot\theta \over \sin\theta}
{{\partial \psi}\over {\partial \tilde \varphi}}
 + (s^{2}\cot^{2}\theta+s)\psi
\nonumber\\
&&
+ 2\left[{sr+ias\cos\theta+(s-1)M}\right] {{\partial \psi}\over
{\partial \tilde t}}
 = 0\, .
\end{eqnarray}

\begin{figure}
\input epsf
\centerline{ \epsfxsize 8.0cm
\epsfbox{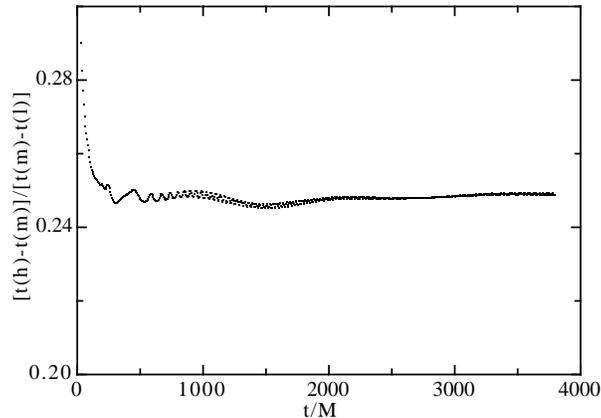}}
\caption{Convergence test for the 2+1 D penetrating Teukolsky code. We 
compute the times at which is field has zeros for three separate resolutions 
[high $t({\rm h})$, medium $t({\rm m})$, and low $t({\rm l})$] and plot
the ratio
$[t({\rm h})-t({\rm m})]/[t({\rm m})-t({\rm l})]$ as a function of time. The 
resolutions used for this test, were $M/40$, $M/20$ and $M/10$ for the high, 
medium and low, respectively. Second order convergence is clear by the value 
of the convergence ratio ($0.25$).  }
\label{ptc_conv}
\end{figure}

For the case of interest, we set $s=0$ in Eq.~(\ref{teukolsky}),  
and include a mass term $\mu^{2}(r^{2}+a^{2}\cos^{2}\theta)\psi$ on the
left hand side of Eq.~(\ref{teukolsky}). 
The PTC implements the numerical integration of the resulting equation by 
decomposing it into azimuthal angular modes and evolving each such 
mode using a reduced 2+1 dimensional linear partial differential equation. 
The results obtained from this code are independent of the choice of 
boundary conditions, because the inner boundary is typically placed inside 
the horizon, whereas the outer boundary is placed far enough that it has 
no effect on the evolution. (As was shown in Ref.~\cite{allen},
close timelike boundaries with Sommerfeld-like boundary conditions do not
allow for the evolution of late-time tails.) Typically, for the
simulations performed in this 
study, the outer boundary was located at $4000M$ and a grid of size 
$40000\times32$ (radial$\times$angle) was used. The initial data was 
always chosen to be a gaussian distribution, centered at $50M$ and with a 
width of $2M$. The PTC is stable and exhibits second order convergence as 
clear from Fig.~\ref{ptc_conv}. 

\begin{figure}
\input epsf
\centerline{ %\epsfxsize 8.0cm
\epsfbox{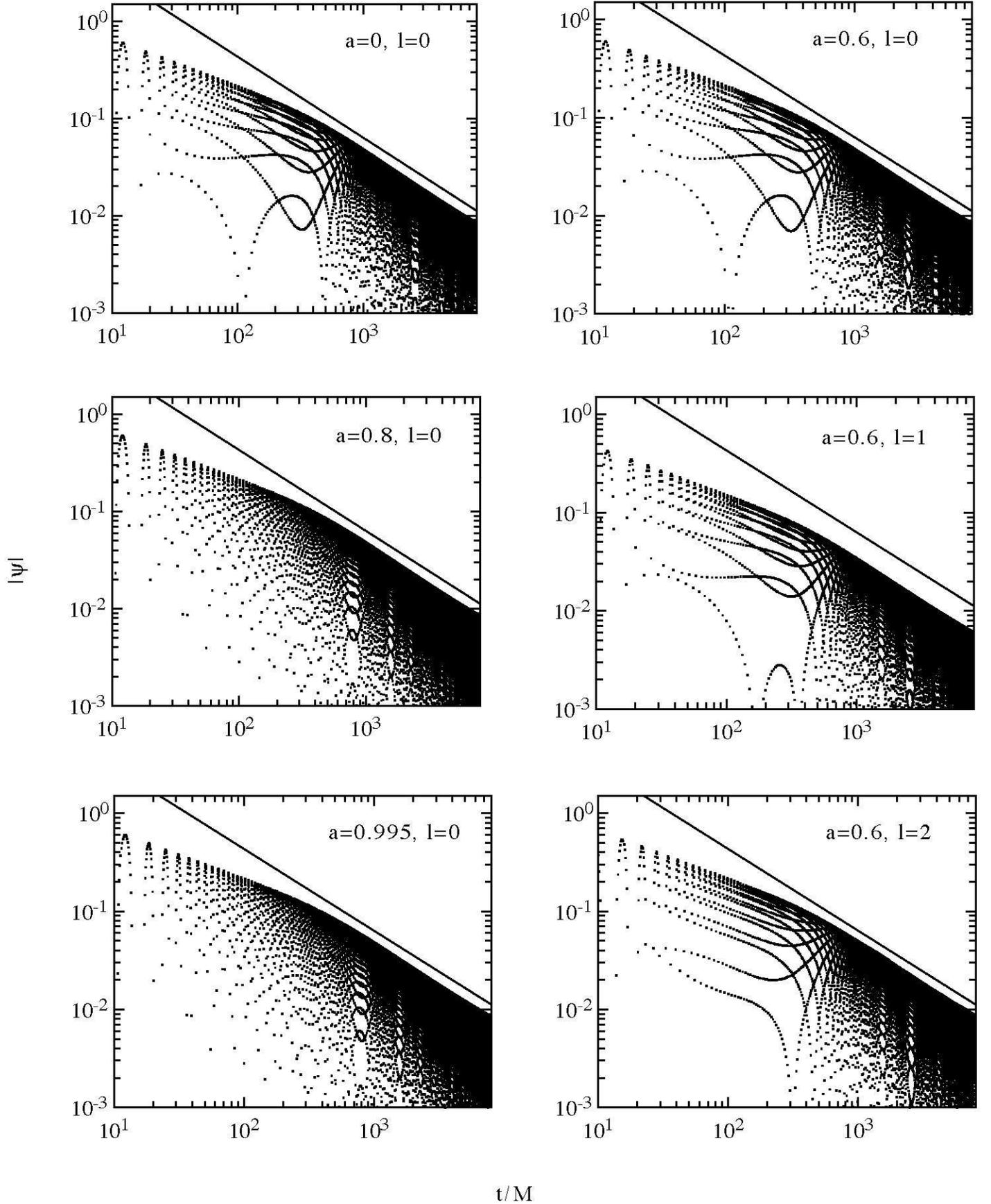}}
\caption{Massive scalar field (with $\mu M=1$) sampled at $r=50M$. The 
initial data are a gaussian distribution, centered at $50M$ and has a 
width of $2M$. The dotted curves show the behavior of the scalar field,
and the solid lines are proportional to $(t/M)^{-5/6}$. The sequence of
plots demonstrate a clear independence of $a/M$, and also of the mode
number $\ell$.}
\label{independence}
\end{figure}

We first demonstrate the independence of the late-time evolution of  
the Kerr parameter ($a/M$) of the background spacetime. Figure 
\ref{independence} (left panels and upper panel on the right) shows tails
from several different evolutions
corresponding to different values of $a/M$, all plotted together. Each 
oscillatory tail shown has the expected period (about $2\pi/\mu$) and a
decay 
rate close to  $t^{-5/6}$. The value of $\mu M$ is chosen here to be $1$. 

We next demonstrate the independence of the oscillatory tail of
the value of $\ell$. Figure \ref{independence} (right panels) shows the
late-time evolution of the massive scalar field ($\mu M=1$) in a Kerr
background spacetime with $a/M=0.6$ for several different values of
$\ell$. There appears to be no deviation from the $t^{-5/6}$ oscillatory
tail for any value of $\ell$.

\begin{figure}
\input epsf
\centerline{ \epsfxsize 8.0cm
\epsfbox{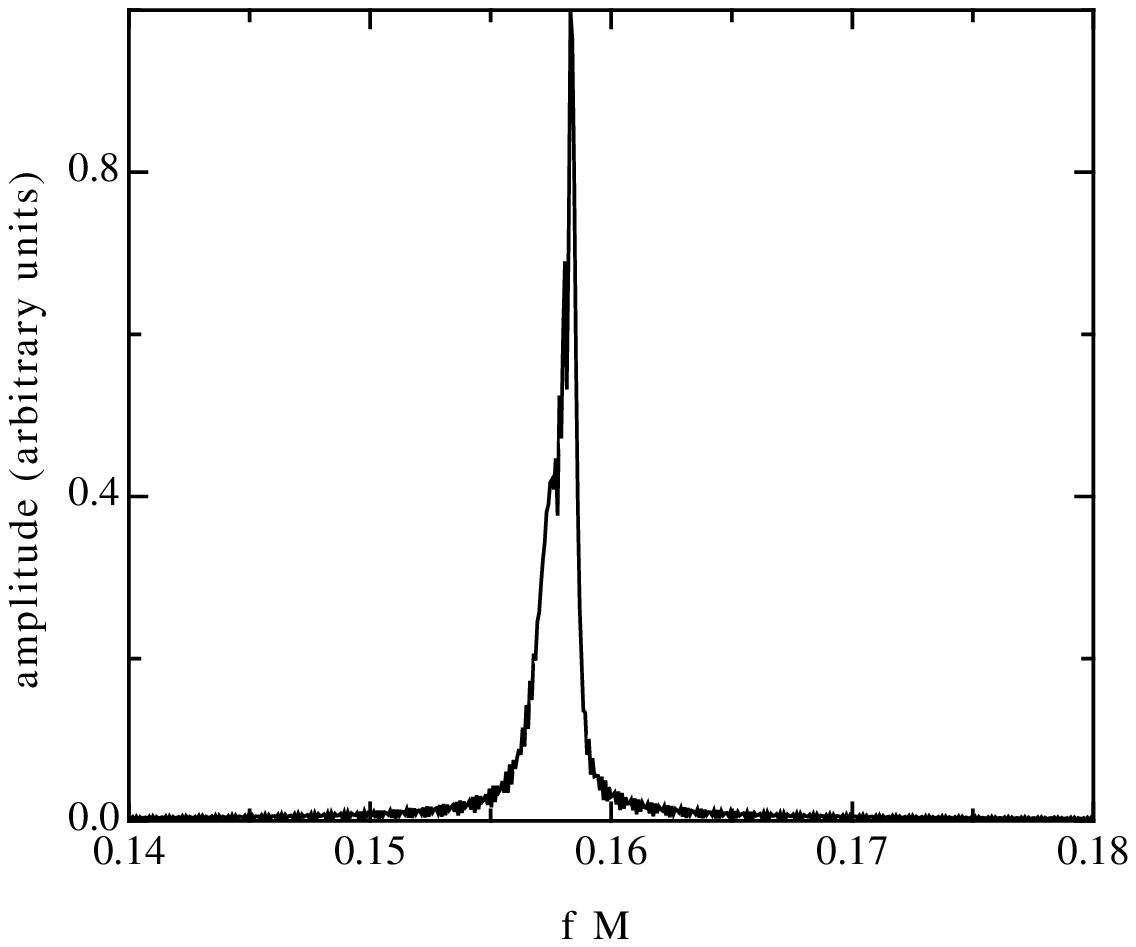}}
\caption{Power spectrum of the field evolving in a Kerr spacetime. 
Here, $\mu M=1$, and $a/M=0.6$. To obtain this plot, the field's values were 
first scaled by $t^{5/6}$ to get close to a pure sinusoid, and then the
Fourier transform was taken.}
\label{ptc_ft}
\end{figure}

\begin{figure}
\input epsf
\centerline{ \epsfxsize 8.0cm
\epsfbox{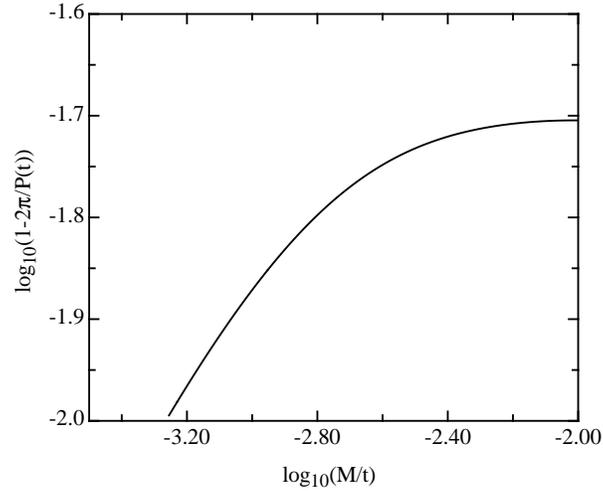}}
\caption{The changing period in Kerr spacetime: $1-2\pi / P(t)$ vs.~$M/t$ 
for the same parameters as in Fig.~\ref{ptc_ft}}
\label{ptc_pt}
\end{figure}

\begin{figure}
\input epsf
\centerline{ \epsfxsize 8.0cm
\epsfbox{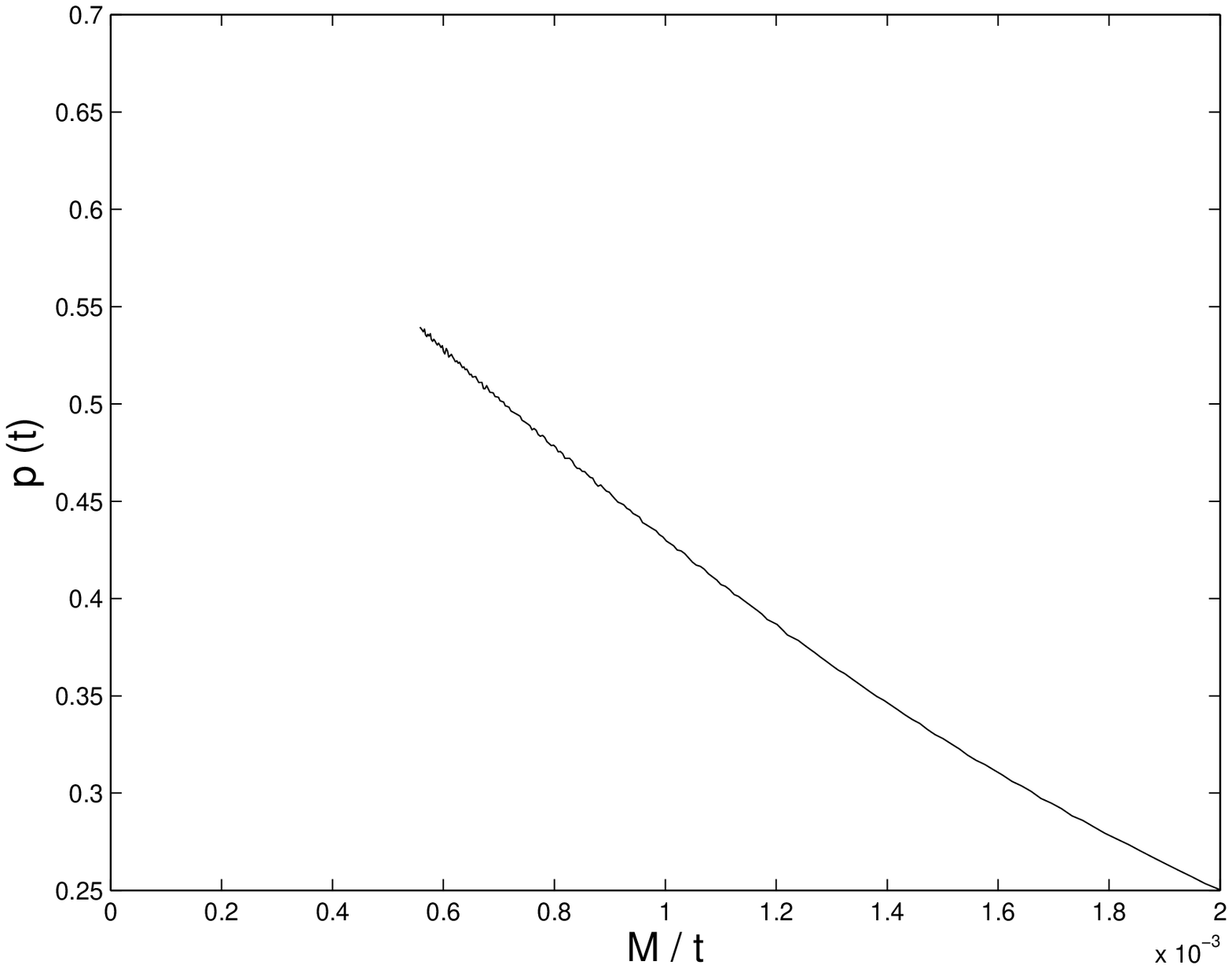}}
\caption{The local slope of the curve in Fig.~\ref{ptc_pt}, $p(t)$, as a
function of $M/t$.}
\label{ptc_dpt}
\end{figure}

Last, we turn to the changing period of the oscillations in the Kerr case. 
In Fig.~\ref{ptc_ft} we plot the power spectrum of the field for the case 
of $a/M=0.6,\, \ell=0$ and $\mu M= 1$. Much like in the Schwarzschild 
case in the previous Section, we observe a one-sided broadening of the
peak  at $(2\pi)^{-1}$. To test whether this indicates a monotonic
increase in 
frequency we plot in Fig.~\ref{ptc_pt}, $1-2\pi/P(t)$ as function of
$t^{-1}$, 
where $P(t)$ is the local period. Indeed, we see in the Kerr case a
decreasing period, with an asymptotic rate (slope of the curve in 
Fig.~\ref{ptc_pt} is shown in Fig.~\ref{ptc_dpt}) of about $0.68\pm
0.02$ which is consistent with the (Schwarzschild) value of $2/3$. 

Thus, in accord with our expectations, the asymptotic late-time behavior
of massive scalar fields in a Kerr spacetime is identical to that in
Schwarzschild background. Specifically, the asymptotic late-time tail is
insensitive to the multipole number $\ell$, and also insensitive to the
spin rate of the black hole. The late-time behavior of massive scalar
fields in black hole spacetimes is universal. There is yet to be an
analytic study in the Kerr context, akin to the one in the Schwarzschild
case \cite{koyama1,koyama2,koyama3}. Also, it should be  noted that all
the numerical simulations reported on in this work were axisymmetric. In a
nonaxisymmetric evolution of massive scalar fields in a Kerr spacetime,
under certain conditions, a very interesting instability arises that has
been studied in the frequency-domain \cite{detweiler}. We are hoping to
return to an extensive study of that case in the time-domain
elsewhere. Lastly, intriguing deviations from the results reported on
here, which appear to introduce a certain low-frequency modulation at
very late times, were reported on in numerical simulations
\cite{burko-unpublished,poisson}. Notably, the analytical analyses do not
show any such phenomena. It remains an open question whether such
phenomena are real physical ones, or a numerical artifact.

\section{Acknowledgements}

The authors thank Jorge Pullin and Steven Detweiler for suggesting a 
study of massive scalar fields in a black hole spacetime, and Hiroko
Koyama and Amos Ori for discussions. GK acknowledges
research support from the University of Massachusetts at Dartmouth and 
also from NSF grant number PHY-0140236. LMB was supported by NSF grant 
PHY-0244605. Some of the numerical simulations were performed at Boston
University's Scientific Computing and Visualization Center. The authors
are grateful for having access to that facility. 

\begin{appendix}

\section{Tails of a massive scalar field in flat spacetime}

In this Appendix we derive the decay rate of Klein-Gordon tails in flat
spacetime. 

The exact Green's function is known \cite{morse}, and is given by
\begin{equation}
g(r,t)=\frac{\delta 
(t-r)}{r}-\frac{\mu J_1[\mu\sqrt{t^2-r^2}]}{\sqrt{t^2-r^2}}\; \Theta (t-r)
\, ,
\end{equation}
for a source at the origin (in space and time). Here, $J_1$ is the Bessel 
function of the first king of order 1. 
This Green's function can
of course be integrated to find the field straightforwardly, but instead
we shall use a more geometric approach to find the field. 

Consider a spherical shell of radius $r_0$, such that the
initial perturbation field is well localized in space and in time. The
only requirement is that the perturbation will be regular everywhere, and
in particular at the location and time of the burst and at the origin
$r=0$. The field $\phi$ satisfies the Klein-Gordon equation 
\begin{equation}
(\Box -\mu^2)\phi=0\, .
\end{equation}
We next define, as usual, $\psi=r\phi$, and write the radial equation for
$\psi$. 

As the field propagates in spherical waves, the field can depend
only on the spacetime interval $s$ from the event of the burst. That is, 
$\psi=\psi (s)$. Inserting this into the radial equation, we find that
$\psi$ satisfies
\begin{equation}
s^2\psi ''+s\psi ' +\mu^2 s^2\psi=0\, ,
\end{equation}
a prime denoting derivative with respect to the interval $s$. This is just
the Bessel equation, with general solution
\begin{equation}
\psi (s)=\tilde{c}_1 J_0(\mu s)+\tilde{c}_2 Y_0(\mu s)\, .
\end{equation}
Here, $J_0$ and $Y_0$ are the Bessel functions of the first and second
kinds, respectively, of order $0$. 
The integration constant $\tilde{c}_2$ must vanish for the field to be
regular at the event of the burst ($s=0$). Therefore, 
\begin{equation}
\psi (s)=\tilde{c}_1 J_0(\mu s)\, .
\end{equation}
For $r={\rm const}$ and for $t\gg r,r_0$, we then find that 
\begin{equation}
\psi \approx c_1\frac{\cos (\mu t-\pi/4)}{(\mu t)^{1/2}}\, .
\end{equation}
This solution does not satisfy the regularity requirement, as the field
$\phi$ diverges at the spatial origin $r=0$. However, the field $\phi$ is
required to be regular at $r=0$, especially at late times ($\mu t\gg 1$). 

\begin{figure}
\input epsf
\centerline{ \epsfxsize 8.0cm
\epsfbox{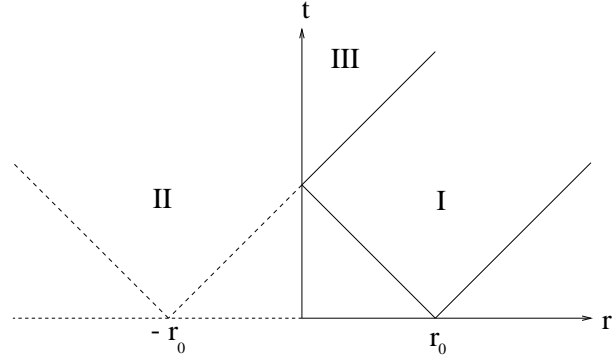}}
\caption{The construction of the ``image charge'' solution. The burst of
the spherical wave is at the event $(0,r_0)$. The domain of dependence of
this event is the union of regions I and III in the diagram. As this
solution is not regular along $r=0$ we introduce an ``image charge,''
i.e., a burst with the opposite sign at the fictitious event $(0,-r_0)$.  
The field due to this image solution in region II is of course unphysical,
but in region III it is, and superposes with the solution of the
original burst in that region.}
\label{image_charge}
\end{figure}

To obtain the physical solution we superpose solutions to the radial
equation, such that the field is regular at the origin. This can be
achieved by an ``image charge.'' The superposition
\begin{equation}
\psi (s)=\tilde{c}_1 \left[ J_0(\mu s_1)-J_0(\mu s_2)\right]\, ,
\end{equation}
is the requested solution, where $s_1^2=t^2-(r-r_0)^2$ and 
$s_2^2=t^2-(r+r_0)^2$. This corresponds to an ``image charge'' at
$r=-r_0$. Obviously, $r=-r_0$ is not a physical point in spacetime. It is
a mathematical way to write a formal solution, where in the physical part
of the domain of dependence of the superposed bursts the solution is the
one sought (see Fig.~\ref{image_charge}). In fact, this is just a
mathematical way to specify regular boundary conditions along $r=0$.

For $r={\rm const}$ and for $t\gg r,r_0$, we then find that 
\begin{equation}
\psi(t\gg r,r_0)\approx c_1\frac{\cos (\mu t-\pi/4)}{(\mu t)^{3/2}}
rr_0\, .
\end{equation}
The phase is unimportant, such that we finally get the result
\begin{equation}
\phi(t\gg r,r_0)\approx c_1r_0\frac{\sin (\mu t)}{(\mu t)^{3/2}}
\, .
\end{equation}
The fall off rate of $t^{-3/2}$ is the one reported on in
Ref.~\cite{morse}.

To find the time dependence of higher $\ell$ modes, we find the derivative
of the spherical solution with respect to the Cartesian coordinate $z$. 
Recall that on the equatorial plane, $\,\partial P_{\ell}(\cos\theta
)/\,\partial z=-(\ell +1)P_{\ell+1}(\cos\theta )/r$. We can thus generate
$\psi_{1}$ from $\psi_0$ by calculating $\,\partial
\psi_{\ell=0}/\,\partial z$, and by the uniqueness of the solution that
should give us the dipole solution
$\psi_{\ell=1}$. Specifically, notice that 
$$\frac{\partial}{\,\partial z}=
\frac{\,\partial r}{\,\partial z}\frac{\,\partial s}{\,\partial r}
\frac{\partial}{\,\partial s}=-\frac{r}{s}\cos\theta\frac{\partial}
{\,\partial s}\, .$$

\noindent Notice the factor $1/s$ in this expression. This implies that
whenever a derivative with respect to $z$ is taken, the exponent of $t$ in
the denominator increases by $1$. (This is the case because of the
sinusoidal function in the numerator. The leading term in $t^{-1}$ comes
from its differentiation, and then the exponent of the denominator does
not change. It is increased by $1$ because of this $1/s$ factor.) 

Carrying $\ell$ times this differentiation with respect to $z$, we find
that 
\begin{equation}
\psi_{\ell}(t\gg r,r_0) \sim \frac{\sin (\mu t)}{t^{\ell+3/2}}\, .
\end{equation}

The reason why the Huygens principle fails in this case is that the phase
velocity  of a plane wave satisfying the Klein-Gordon equation is
\begin{equation}
v_{\rm ph}=\frac{1}{\sqrt{1-(\mu /\omega )^2}}\, ,
\end{equation} 
so that whenever $\mu\ne 0$, different frequencies which make up a wave
packet travel at different speeds. As noted in \cite{morse}, it is not
surprising that the associated plane waves do not arrive at the
observation point with the same relative phase they had at the
beginning. In fact, spacetime behaves like a dispersive medium for the
Klein-Gordon equation, as the wave number $k=\omega\sqrt{1-(\mu /\omega
)^2}$ is no longer a linear function of $\omega$.

\end{appendix}

\end{document}